\documentclass[prx,twocolumn,showpacs,superscriptaddress]{revtex4}
\usepackage{amsmath}
\usepackage{amssymb}
\usepackage{amsthm}
\usepackage{amsfonts}
\usepackage{listings}
\usepackage{enumerate}
\usepackage{latexsym}
\usepackage{color}
\usepackage{multirow}
\usepackage{makecell}
\usepackage{bm}
\usepackage{hyperref}
\hypersetup{
 pdfnewwindow=true, colorlinks=true,
 linkcolor=blue, anchorcolor=blue,
 citecolor=blue, filecolor=blue,
 menucolor=blue, urlcolor=blue}

\usepackage{psfrag}
\usepackage{mathrsfs}
\usepackage{bm}
\usepackage{graphicx}
\usepackage{subfigure}

\RequirePackage[normalem]{ulem} 
\RequirePackage{color}\definecolor{RED}{rgb}{1,0,0}\definecolor{BLUE}{rgb}{0,0,1} 

\newcommand{\beq}{\begin{equation}}
\newcommand{\eneq}{\end{equation}}

\def\bk{{\bf k}}

\def\kmnaso{K$_{2}$Mn$_3$(AsO$_{4}$)$_3$}

\def\ie{{\it i.e.},\ }

\begin{document}

\tolerance 10000

\newcommand{\vk}{{\bf k}}

\draft

\title{Magnetic Weyl semimetal in \kmnaso~with the minimum number of Weyl points}

\author{Simin Nie}
\thanks{These authors contributed equally}
\affiliation{Department of Materials Science and Engineering, Stanford University, Stanford, California 94305, USA}

\author{Tatsuki Hashimoto}
\thanks{These authors contributed equally}
\affiliation{Department of Mechanical Engineering, Stanford University, Stanford 94305, USA}

\author{Fritz B. Prinz}
\email{fprinz@stanford.edu}
\affiliation{Department of Materials Science and Engineering, Stanford University, Stanford, California 94305, USA}
\affiliation{Department of Mechanical Engineering, Stanford University, Stanford 94305, USA}
\date{\today}

\begin{abstract}
The ``Hydrogen atom" of magnetic Weyl semimetals, with the minimum number of Weyl 
points, have received growing attention recently due to the possible 
presence of Weyl-related phenomena. Here, we report a nontrivial electronic 
structure of the ferromagnetic alluaudite-type compound \kmnaso. It exhibits only a 
pair of Weyl points constrained in the $z$-direction by the two-fold rotation symmetry, 
leading to extremely long Fermi arc surface states. In addition, the study of its 
low-energy effective model results in the discovery of various topological 
superconducting states, such as  the ``hydrogen atom" of a Weyl superconductor. 
Our work provides a feasible platform to explore 
the intrinsic properties related to Weyl points, and the related device applications.
\end{abstract}

\maketitle

%
%
\emph{Introduction.---}
The realization of elementary particles (\ie 
Dirac, Weyl and Majorana fermions) in condensed matter has received growing attention due 
to high scientific interest and promising applications in novel quantum 
devices~\cite{wang2012dirac,wang2013three,murakami2007phase,wan2011topological,burkov2011weyl,xuchern,liu2014,weng2015weyl,burkov2016topological,nie2017topological,yan2017topological,science2,science1,science3,sato2017topological,armitage2018weyl,lv2021,zhang2018observation,nadj2014observation,wang2018evidence,nie2018}. 
Compared with Dirac and Majorana fermions, Weyl fermions do not need any specific symmetry
protection (but the lattice translation symmetry) to guarantee their existence. Weyl semimetals
exhibit linear dispersion around discrete doubly degenerate points [termed Weyl points (WPs)], 
whose low-energy excitation exactly satisfies the Weyl equation of quantum field theory~\cite{weyl1929elektron}. In 
momentum space, the WPs with positive and negative chirality can be viewed as the ``source" and
``drain" points of the ``magnetic field"~\cite{fang2003anomalous}, respectively. According to the
``no-go theorem"~\cite{nielsen1983adler}, the total chirality in the entire three-dimensional 
Brillouin zone (BZ) must be zero, \ie the WPs always appear in pairs of opposite chirality. 
Therefore, the minimum numbers of WPs in nonmagnetic and magnetic Weyl semimetals are four and
two, respectively. This type of Weyl semimetals, called ``hydrogen atom" of Weyl semimetals
\cite{bernevig2015s}, are of great interest due to the simple phenomena related 
only to WPs, such as large negative magnetoresistance~\cite{huang2015prx,anomaly,arnold2016negative} 
and large anomalous Hall conductivity~\cite{burkov2014anomalous,weng2015quantum,liu2018giant,wang2018large}. These properties 
are particularly important for related device designs.

Recently, significant progress in symmetry-based strategies
\cite{po2017symmetry,bradlyn2017topological,khalaf2018symmetry,kruthoff2017topological,song2018quantitative,tang2019efficient} 
greatly accelerates the discovery of both topological insulating states and topological 
semimetals~\cite{vergniory2019complete,tang2019comprehensive,zhang2019catalogue}. However, these
elegant strategies can lead to ``false-negative" results in identifying Weyl 
semimetals~\cite{Qian2020,gao2021high}. Due to the inexistence of a suitable topological invariant 
characterizing the Weyl semimetals, the search of them is comparatively difficult. Surprisingly, 
there is a special system, \ie magnetic centrosymmetric system, for which the topological invariant
$\chi$ can be defined by 
\begin{equation}
(-1)^\chi\equiv\prod_{j=\{1,2,\cdots,n_{occ}\},~\Gamma_i=\text{TRIMs}} \xi^j_{i}, 
\label{chi}
\end{equation}
where $\xi_i^j$ is the parity eigenvalue ($\pm1$) of the $j$-th band at the
time-reversal-invariant-momentum (TRIM) $\Gamma_i$, and $n_{occ}$ is the total number of the 
valence bands. If $\chi=1$, the system must have band crossing points around the Fermi level, and 
may be a Weyl semimetal~\cite{zjwang,hughes,eub6}. Although the ``hydrogen atom" of nonmagnetic Weyl 
semimetal has been discovered~\cite{belopolski2017signatures}, and there are some high-throughput 
screening methods for Weyl semimetals~\cite{Ivanov2019,xu2020high,xu2020comprehensive,gao2021high},
the discovery of ``hydrogen atom" of magnetic Weyl semimetals is still challenging but 
represents a highly desirable state.

\begin{figure*}[ht]
\includegraphics[width=6.2in]{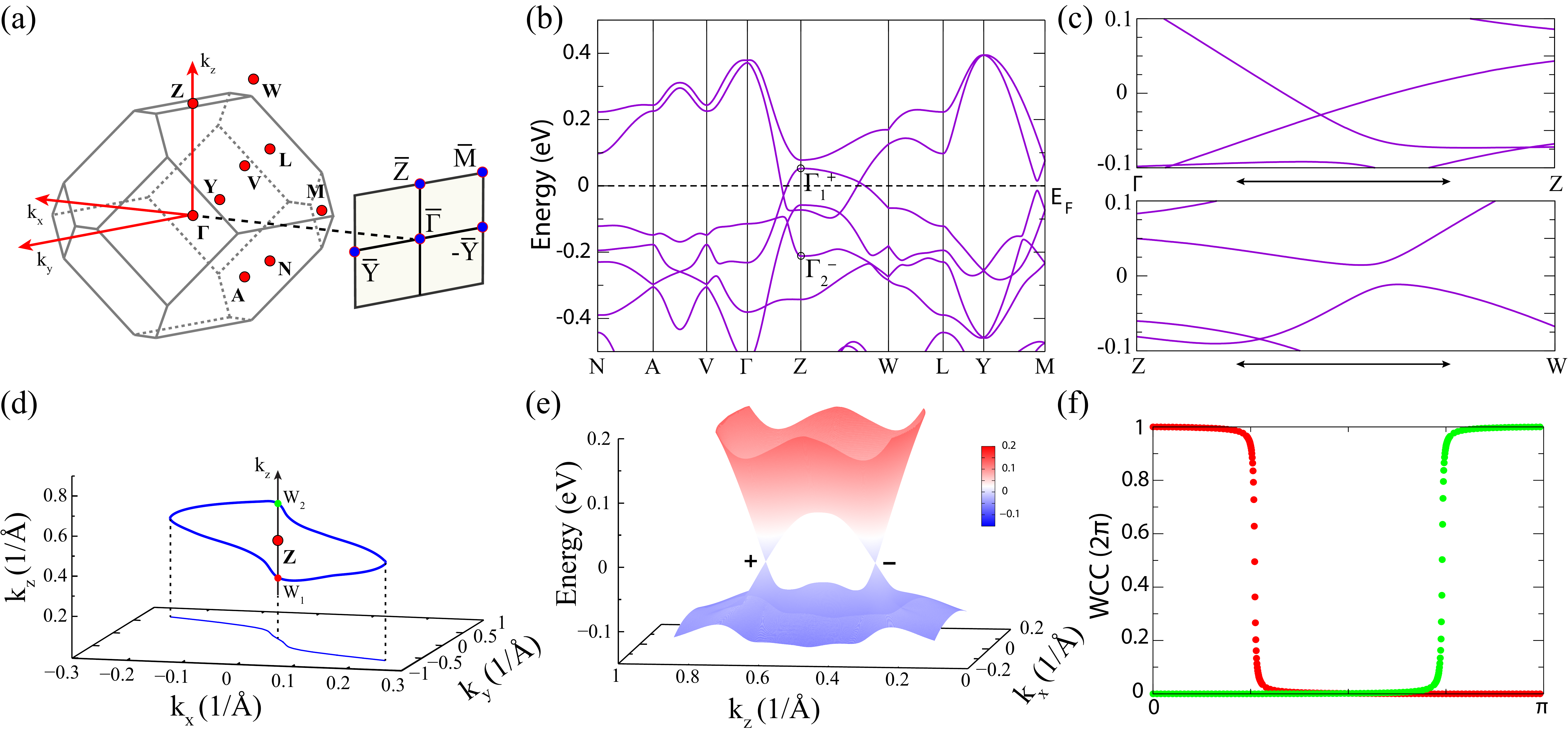}
      \caption{(color online). Electronic structure of \kmnaso, (a) The bulk BZ and the projected $yoz$-surface BZ with high-symmetry points. (b) The GGA+U band structure of \kmnaso. (c) The GGA+U+SOC band structure around the band crossing points near the Fermi level. (d) The nodal line around the Z point in the BZ. (e) Three-dimensional band structure around the WPs. (f) The evolution of Wannier charge centers (WCCs) on two spheres enclosing W$_1$ (red dots) and W$_2$ (green dots), respectively. 
      The coordinates of W$_1$ and W$_2$ are $\vec{K}^Z\mp(0,0,0.156621)$ in units of $1/\text{\AA}$, respectively.}
\label{fig1}
\end{figure*}

In this work, the topological properties of alluaudite-type compound \kmnaso~are systematically
studied based on first-principles calculations and low-energy effective model analysis. The total
energy calculations show that \kmnaso~favors a ferromagnetic (FM) ground state with magnetic 
momentum in the $z$-direction (\ie [$\bar{1}$01]-direction). The FM \kmnaso~hosts only a pair of 
WPs with opposite chirality around the $Z$ point, which are constrained in the $z$-direction by 
the two-fold rotation symmetry around the direction. More interestingly,
extremely long Fermi arcs exist on the $yoz$ plane of \kmnaso, which can be easily observed by
angle-resolved photoemission spectroscopy experiment. Based on its low-energy effective model, 
the possible nontrivial superconductor (SC) states are explored, which gives rise to the discovery 
of various novel states, such as ``hydrogen atom" of Weyl SC. ``Hydrogen atom" of FM
Weyl semimetal in \kmnaso~greatly facilitates the study of Weyl-related physics only and the 
device applications of Weyl semimetals.

\emph{Crystal and magnetic structures of \kmnaso.---}
\kmnaso~has been experimentally fabricated in bulk since 2012~\cite{chaalia2012k}. The crystal structure
was identified to be of the alluaudite-type which can be described 
by the formula A(2)A(1)M(1)M(2)$_2$(XO$_4$)$_3$
\cite{moore1971crystal}. 
This structure contains two sets of tunnels in the [010]-direction with A atoms 
at the centers, embedded in the M(1)M(2)$_2$(XO$_4$)$_3$ framework (see details in Section B of the 
Supplemental Material~\cite{supp}). The framework is formed by chains of edge-sharing MO$_6$ octahedra, which 
are linked together by the XO$_4$ tetrahedra. The remarkable flexibility of the framework allows 
cation substitution in the X and M sties and tolerates a wide range of compositional variations,
leading to the presence of interesting electrical-transport and magnetic properties
~\cite{kim2017development,karegeya2017one,chouaibi2001neutron,dwibedi20152,essehli2011crystal}.
Given the partial occupation of Mn $d$-orbitals and the observation of magnetic property
in alluaudite-type manganese sulphate~\cite{dwibedi20152}, we anticipate a possible magnetic
ground state in \kmnaso. By using the generalized gradient approximation (GGA)+Hubbard-U (GGA+U)
method, the possible magnetic structures of \kmnaso~have been explored (see calculation method in
Section A of the Supplemental Material~\cite{supp}). Here, we studied eleven collinear magnetic
configurations, including ten antiferromagnetic (AFM)-like configurations and one FM configuration.
The calculations show that the total energy of the nonmagnetic state is about 45 eV/unit cell 
lower than that of the magnetic states. Moreover, the FM configuration lowers the total
energy by dozens of meV compared to AFM-like configurations, and is the ground
state of \kmnaso~(see details in Section C of the Supplemental Material~\cite{supp}).

\begin{table}[t]
\caption{The numbers of even and odd valence bands at eight TRIMs. The positions of the TRIMs are given in three primitive reciprocal vectors.}
\begin{tabular}{ p{1.5cm}|p{2cm}|p{2cm}|p{2cm}}
\hline
\hline
\hfil TRIM & \hfil Position & \hfil Even parity & \hfil Odd parity\\
\hline
\hfil $\Gamma$ &\hfil (0,0,0) & \hfil 122 & \hfil 130 \\
\hline
\hfil Y                &\hfil (0,0.5,0) &\hfil 126  &\hfil 126 \\
\hline
\hfil Z                &\hfil (-0.5,0,0.5) &\hfil 121  &\hfil 131 \\
\hline
\hfil A                &\hfil (0.5,0,0) &\hfil 130  &\hfil 122 \\
\hline
\hfil V               &\hfil (0,0,0.5) &\hfil 130  &\hfil 122 \\
\hline
\hfil L                &\hfil (0,0.5,0.5) &\hfil 126  &\hfil 126 \\
\hline
\hfil N                &\hfil (0.5,0.5,0) &\hfil 126  &\hfil 126 \\
\hline
\hfil M                &\hfil (0.5,0.5,0.5) &\hfil 126  &\hfil 126 \\
\hline
\hline
\end{tabular}
\label{parity}
\end{table}

\emph{Band structures of \kmnaso.---}
Based on the FM structure, the band structures of \kmnaso~are calculated and shown 
in Fig. \ref{fig1}. In the GGA+U band structure without the consideration of spin-orbit coupling 
(SOC), there is a band inversion between the $\Gamma_1^+$ band and the $\Gamma_2^-$ band at the 
$Z$ point, which gives rise to a nodal line circled around the point, as shown in Figs. \ref{fig1}(b)
and \ref{fig1}(d). The nodal line is protected by the coexistence of inversion symmetry and time
reversal symmetry (TRS). When SOC is included, the overall shape of the band structure around the
Fermi level (E$_F$) changes very little except the band gap opening at the nodal line (Fig. 
\ref{fig1}(c)). As the system still has the inversion symmetry, and the irreducible representations
of the two inverted bands become $\Gamma_3^+$ and $\Gamma_4^-$, respectively, the
topological invariant $\chi$ in Eq. (\ref{chi}) is well defined. According to the numbers of even 
and odd valence bands at eight TRIMs in Table \ref{parity}, $\chi$ is computed to be 1, guaranteeing
the presence of band crossing points around the Fermi level. 
Our calculations show that band gaps open up along the nodal line except two gapless 
points (\ie WPs) on the $k_z$-axis, as shown in Figs. \ref{fig1}(c) and \ref{fig1}(e). The WPs with opposite 
chirality (Fig. \ref{fig1}(f)) are constrained on the $k_z$-axis due to the two-fold rotation symmetry 
$\{\hat{C}_2^z|0,\frac{1}{2},0\}$.

\begin{figure}[!t]
\includegraphics[width=3.4in,angle=0]{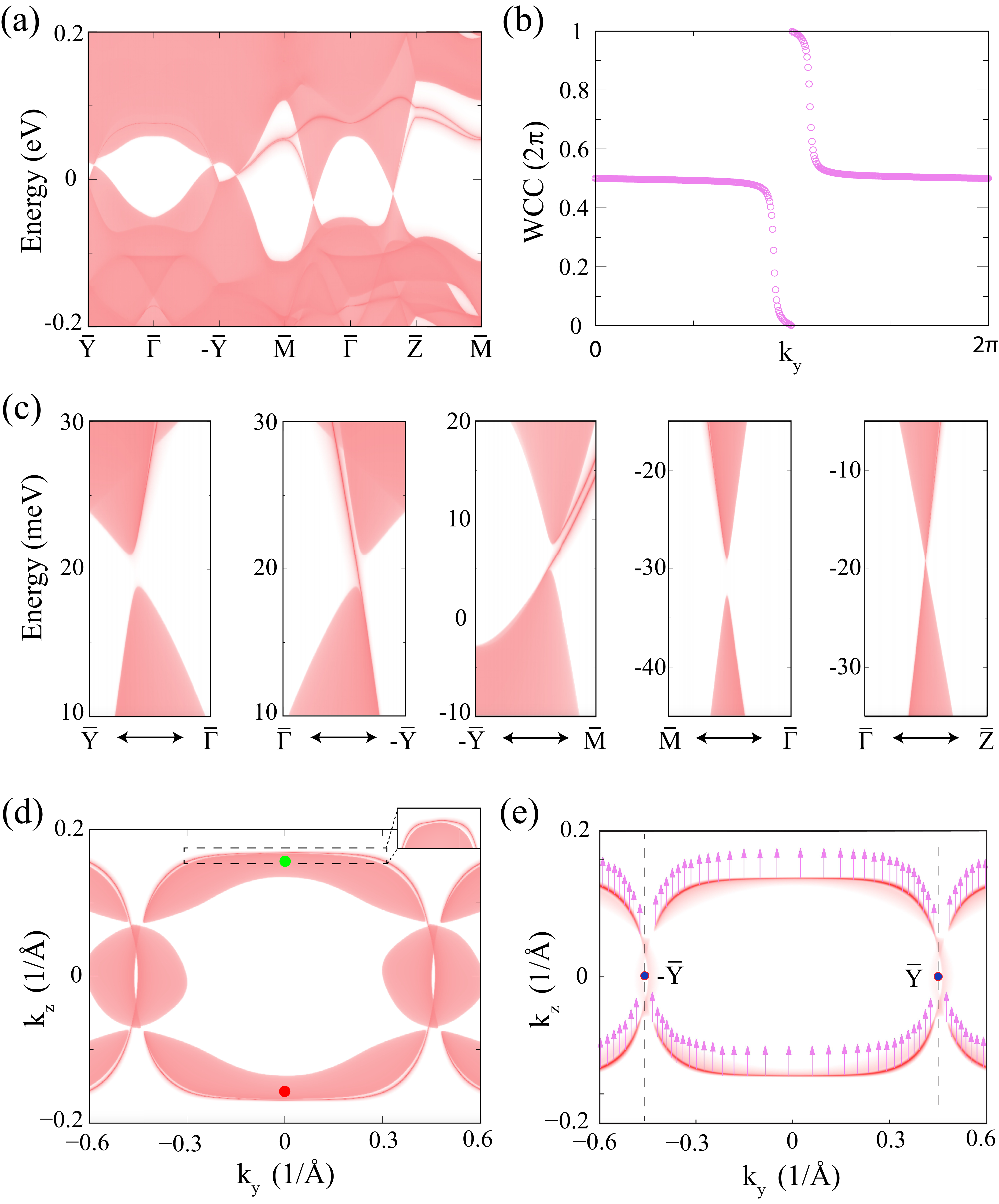}
      \caption{(color online). Surface states and Fermi arcs. (a,c) Energy and momentum dependence of the LDOS on $yoz$ surface for \kmnaso. (b) The evolution of WCCs as a function of $k_y$ for the $k_z=0$ plane. (d) Fermi-arc surface states on $yoz$ surface at Fermi level. 
      The projected WPs are shown as red and green dots for different chirality.
      It is worth noting that the surface projection of the $Z$ point is the same as the $\bar{\Gamma}$ point as 
they differ by a primitive reciprocal vector of the surface BZ (see the definitions of the 
surface primitive reciprocal vectors in Supplemental Material). So, the projected WPs 
shown as red and green dots for different chirality are located at $(k_y,k_z)=(0,\mp0.156621)$ in units of $1/\text{\AA}$.
      (e) same as (d) but with the Fermi arc states at Fermi level highlighted. The spin-texture 
of the Fermi arc states are represented by arrows.}
\label{fig2}
\end{figure}

\emph{Topological surface states and Fermi arcs.---}
In view of the fact that one hallmark of Weyl semimetal is the existence of Fermi-arc surface
states, maximally localized Wannier functions (MLWFs) for the $d$ orbitals of Mn and $p$ orbitals 
of O are constructed, which are used to build the Green's functions of the semi-infinite slabs 
by using an iterative method. The local density of states (LDOS) on $yoz$ surface, extracted 
from the imaginary parts of the surface Green's functions, are shown in Fig. \ref{fig2}. On the
$yoz$ surface, the WPs are projected to the $\bar{\Gamma}$-$\bar{Z}$ direction, leading to the
existence of gapless dispersion in the direction, as shown in Figs. \ref{fig2}(a) and \ref{fig2}(c). Along the
$\bar{\text{Y}}$-$\bar{\Gamma}$-$(-\bar{\text{Y}})$ line, there is one surface state crossing the Fermi
level, which is consistent with the nontrivial Chern number (\ie $C=-1$) of the $k_z=0$ plane, as
shown in Figs. \ref{fig2}(b) and \ref{fig2}(c). In addition, the constant-energy contour of the
surface states clearly shows that two Fermi-arc surface states derived from two bulk electron
pockets in the $(-\bar{Z})$-$\bar{\Gamma}$-$\bar{Z}$ direction (enclosing two opposite-chirality WPs,
respectively) are buried in the same bulk hole pocket in the $\bar{\text{Y}}$-$\bar{\Gamma}$-$(-\bar{\text{Y}})$ direction, 
as shown in Figs. \ref{fig2}(d) and \ref{fig2}(e). 
Because the Fermi arc surface states are slightly buried in the bulk states at Fermi level, the Fermi arc states 
around $\bar{\text{Y}}$ or $-\bar{\text{Y}}$ are blurred. However, a Fermi arc line crossing the $k_z=0$ line can still be seen.
Compared with Fermi-arc surface states in
well-known nonmagnetic Weyl semimetals (such as TaAs~\cite{weng2015weyl}), there are two distinct
features of the Fermi arcs in \kmnaso: (i) the Fermi arcs are extremely long, which are desirable 
to the related device applications; (ii) the two states at $\bar{k}$ and $-\bar{k}$ on the
constant-energy contour carry parallel spin alignment.

\emph{Low-energy effective model.---}
In order to understand the main feature of the low-energy band structure of \kmnaso, an effective
low-energy $2\times 2$ $k\cdot p$ model is constructed (see details in Section D of the Supplemental
Material~\cite{supp}). When SOC is ignored and the constraints placed by all symmetries [including inversion
symmetry $\hat{I}$, glide mirror symmetry $\hat{g}_z=\{\hat{M}_z|0,\frac{1}{2},0\}$ (the translation
is given in units of three primitive lattice vectors) and TRS $\mathcal{T}$] are considered, the 
model with the $\Gamma_1^+$ band and the $\Gamma_2^-$ band as the basis can be up to second order 
of $\mathbf{k}$ written as 
\begin{eqnarray}
&H^{Z}(\mathbf{k})= d_{y}(\mathbf{k})\sigma_{y}+d_{z}(\mathbf{k})\sigma_{z}, \label{kp1} \\
&\text{with } \bk\equiv (k_x,k_y,k_z)=(K_x,K_y,K_z)-\vec K^Z, \nonumber
\end{eqnarray}
where $\mathbf{k}$ is the momentum vector relative to the $Z$ point; $\sigma_{y,z}$ are Pauli
matrices. $d_{y}(\mathbf{k})$ and $d_{z}(\mathbf{k})$ are odd and even real functions of
$\mathbf{k}$, respectively. The eigenvalues of Eq. (\ref{kp1}) are 
$E({\mathbf{k}})=\pm\sqrt{d_y^2({\mathbf{k}})+d_z^2({\mathbf{k}})}$. 
The degenerate band crossings require 
\begin{eqnarray}
d_{y}(\mathbf{k})&=&b_{2}k_{x}+b_{3}k_{y}=0,  \label{eq3}\\
d_{z}(\mathbf{k})&=&c_{1}+c_{4}k_{x}^{2}+c_{5}k_{y}^{2}+c_{6}k_{z}^{2}+c_{7}k_{x}k_{y}=0.
\label{eq4}
\end{eqnarray}
By substituting Eq. (\ref{eq3}) into Eq. (\ref{eq4}), it is easy to get the following equality
 \begin{eqnarray}
-\frac{(b_{3}^2 c_4+b_2^2c_{5}-b_2b_3c_7)}{b_2^2c_1}k_{y}^{2}-\frac{c_{6}}{c_1}k_{z}^{2}=1.
\label{eq5}
\end{eqnarray}
As the two bands are inverted along the $\Gamma$-$Z$ direction, we can get the requirement 
$c_1 \cdot c_6  <0$. Therefore, the prefactor of $k_z^2$ (\ie $-c_{6}/c_1$) is greater than 0, 
and the Eq. (\ref{eq5}) is a hyperbola or an ellipse depending on the sign of the prefactor of
$k_y^2$. By fitting the first-principles band structure with the model, the values of these
parameters are obtained, as shown in Table S4. It is easy to find the sign of the prefactor 
of $k_y^2$ is plus. Therefore, the analysis of the model shows that the band crossing points 
around the Fermi level in \kmnaso~form an ellipse around the $Z$ point, which is consistent with 
our first-principles calculations.

After the consideration of SOC, the TRS is broken, while the glide mirror symmetry and inversion
symmetry are preserved. Therefore, it is easy to get the effective model with SOC, as shown below
\begin{eqnarray}
&H^{Z}_{SOC}(\mathbf{k})= d_{x}(\mathbf{k})\sigma_{x}+d_{y}(\mathbf{k})\sigma_{y}+d_{z}(\mathbf{k})\sigma_{z}, \label{kpsoc} 
\end{eqnarray}
where $d_{x}(\mathbf{k})$ is an odd real function of $\mathbf{k}$ (see details in Section D of the
Supplemental Material~\cite{supp}). The eigenvalues of Eq. (\ref{kpsoc}) are
$E({\mathbf{k}})=\pm\sqrt{d_x^2({\mathbf{k}})+d_y^2({\mathbf{k}})+d_z^2({\mathbf{k}})}$. 
The degenerate band crossings require $d_{x}(\mathbf{k})=d_{y}(\mathbf{k})=d_{z}(\mathbf{k})=0$. 
Because $c_1 \cdot c_6  <0$, there are two gapless points at (0, 0, $\pm\sqrt{-c_1/c_6}$), which 
are WPs in the $(-Z)$-$\Gamma$-$Z$ direction. Therefore, the $k\cdot p$ models can capture the
low-energy physics in \kmnaso.

\begin{figure}[!t]
\includegraphics[width=3.4in,angle=0]{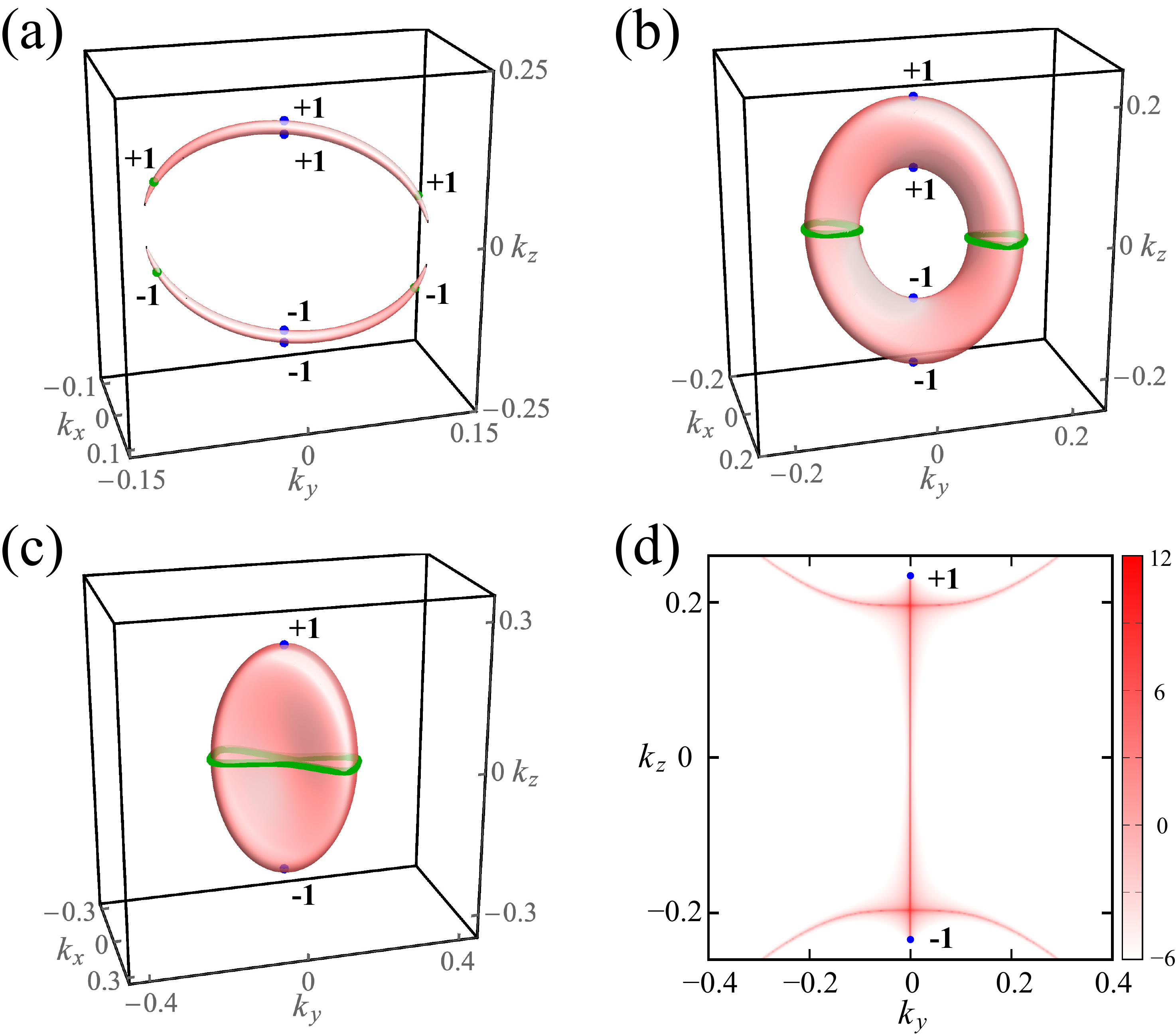}
      \caption{(color online). The FS and the superconducting gap structure. The crescent (a), torus (b)
       and ellipsoid-like (c) FS of \kmnaso~with chemical potential at $\mu=0.01$, $0.05$ and $0.1$ eV, respectively.
       The gapless structures of the pairing potential $\Delta_{Au}$ (green dots or lines) and $\Delta_{Bu}$ (blue dots).
       (d) The crossed surface Andreev bound states in the $B_u$ state with $\mu=0.1$ eV. The $Z$ or $\bar{Z}$ point is 
       chosen as the origin point. The charges of the WPs are indicated.
}
\label{fig3}
\end{figure}

\emph{The nontrivial SC states.---}
Weyl semimetals without TRS are considered as a promising platform to realize the nontrivial SC
states with Majorana quasi-particles \cite{Sato_2017}. This is because, in the presence of the
inversion symmetry but absence of the TRS, the spin configuration of the Cooper pair is parallel,
namely, conventional spin-singlet $s$-wave state is excluded from the candidate of the pairing
symmetry. Next, we briefly discuss the possible SC state in \kmnaso~using the $ k\cdot p$
model derived above. The remarkable feature of \kmnaso~is that the shape of the
Fermi surface (FS) can be easily tuned by doping as its structure allows a wide range of cationic
substitution in K and Mn sites. We hence also clarify how the SC properties develop according to  
the evolution of the FS.

{\renewcommand\arraystretch{1.5}
\begin{table}[t]
\begin{center}
\caption{SC gap structures and the type of SC of the possible pairing states. 
The number and position of the gapless structure are shown in bracket.}
\begin{tabular}{c|c|c|c}
\hline\hline
Rep.&crescent FS&torus or ellipsoid FS&type of SC\\
\hline\hline
$\Delta_{A_g}$&Nodal surface&Nodal surface&gapless\\
\hline
$\Delta_{B_g}$&Nodal surface&Nodal surface&gapless\\
\hline
$\Delta_{A_u}$&\makecell[c]{WPs \\(4, generic)}& \makecell[c]{Nodal line \\ (2 or 1, $k_z$=0 plane)}&$p_z$+chiral-$f$\\
\hline
$\Delta_{B_u}$&\makecell[c]{WPs \\(4, $k_z$-axis)}&\makecell[c]{WPs \\(4 or 2, $k_z$-axis)}&chiral-$p$\\
\hline\hline
\end{tabular}
\label{tab_SC_gap}
\end{center}
\end{table}}

To describe the SC state, we start from the mean-field Hamiltonian in the Bogoliubov-de Gennes (BdG)
formalism:
\begin{align}
   H_{\rm BdG}&=\int{d{\bf k}{\bm c}_{\bf k}^\dagger H({\bf k}) {\bm c}_{\bf k}},\\
   H({\bf k})&=
    \begin{pmatrix}
     H^Z_{SOC}({\bf k})-\mu&\Delta({\bf k})\\
    \Delta^\dagger({\bf k})&-H^{Z*}_{SOC}(-{\bf k})+\mu\\
    \end{pmatrix},\\
    {\bm c}_{\bf k}&=(c_{1,{\bf k}},c_{2,{\bf k}},c_{1,-{\bf k}}^\dagger,c_{2,-{\bf k}}^\dagger),
\end{align}
where $\Delta({\bf k})$ is the SC pairing potential and $\mu$ is the chemical potential. For the above BdG Hamiltonian, we consider 
the following pairing potentials which can be classified into four irreducible representations of the $\mathcal{C}_{2h}$ point group:
    $\Delta_{A_g}\equiv i \Delta_0 k_x \sigma_x$,
    $\Delta_{B_g}\equiv i \Delta_0 k_z \sigma_x$,
    $\Delta_{A_u}\equiv i \Delta_0 k_z \sigma_0$,
    $\Delta_{B_u}\equiv i \Delta_0 \sigma_y$.
To clearly see the SC gap structure of each pairing state, we derive the single band representation
of the pairing potential (see details in Section E of the Supplemental material~\cite{supp}).

It is found that the conduction band components of the even-parity pairings, $\Delta_{A_g}$ and
$\Delta_{B_g}$, are zero, which means that they have similar nodal surfaces (\ie crescent, torus 
and ellipsoid-like nodal surfaces). Interestingly, the crescent nodal surface states are topologically
protected and characterized by a pair of topological charges (\ie $\mathbb{Z}_2$ $\oplus$
2$\mathbb{Z}$) (see details in Section F of the Supplemental material~\cite{supp})~\cite{nodalsurf1,nodalsurf2}. 
However, the even-parity pairing states are less likely to occur
due to the inexistence of the superconducting gap.
On the other hand, for the odd-parity pairings, the
single band representations are obtained as:
\begin{align}
    \tilde{\Delta}_{A_u}^c&=\frac{ik_z\Delta_0}{2}
    \left[
    1+\frac{d_z({\mathbf{k}})}{\varepsilon}
    -
    \left(
    1-\frac{d_z({\mathbf{k}})}{\varepsilon}
    \right)
    e^{-2i\theta}
    \right]
    ,\label{eq_DcAu}\\
    \tilde{\Delta}_{B_u}^c&=-\frac{\Delta_0}{\varepsilon}(d_x({\mathbf{k}})-id_y({\mathbf{k}}))\label{eq_DcBu},
\end{align}
indicating that $\Delta_{A_u}$ and $\Delta_{B_u}$ can be effectively considered as $p_z$+chiral
$f$-wave state and chiral $p$-wave state, respectively. By solving $\tilde{\Delta}_{\alpha}^c=0$
($\alpha=A_u$ or $B_u$), it is found that $\Delta_{A_u}$ and $\Delta_{B_u}$ pairings exhibit WPs 
or nodal lines depending on the position of the chemical potential $\mu$, as shown in Fig. 
\ref{fig3}. When the FS is the crescent shape ($\mu=0.01$ eV), WPs on the generic momenta and $k_z$ axis 
are presented in the $\Delta_{A_u}$ and $\Delta_{B_u}$ pairings, respectively. On the 
other hand, when the FS is a torus ($\mu=0.05$ eV) or an ellipsoid-like ($\mu=0.1$ eV) surface, the 
nodal lines (WPs) appears on the $k_z=0$ plane ($k_z$ axis) for the $\Delta_{A_u}$ ($\Delta_{B_u}$)
pairing. The SC gap structures of the possible pairing states are also summarized in Table
\ref{tab_SC_gap}. 
When the chemical potential is low, there are two pairs of WPs on the $k_z$ axis for the $B_u$
state. As the chemical potential increases, one of the pairs approach each other. They meet 
and annihilate at the BZ's center finally (see details in Section F of the Supplemental 
Material). Therefore, there is only a pair of WPs left, \ie it is a ``hydrogen atom" of Weyl SC.
The WPs lead to novel crossed surface Andreev bound states, which are associated with the
nontrivial Chern number~\cite{PhysRevLett.114.096804}, as shown in Fig. \ref{fig3}(d).

\emph{Conclusion.---}
In summary, we propose that ``hydrogen atom" of magnetic Weyl semimetal can be realized in the FM alluaudite-type 
compound \kmnaso. The presence of two WPs constrained in the z-direction leads to extremely long Fermi arcs on the $yoz$ plane, 
which are expected to be easily observed in measurements. Moreover, we have shown that the odd-parity chiral $p$-wave and chiral
$f$-wave state can be realized in \kmnaso. Depending on the shape of the FS, a wide variety of the superconducting gap structure 
can be realized in the odd-parity SC, such as ``hydrogen atom" of Weyl SC with novel crossed surface Andreev bound states.

\emph{Acknowledgments.---}
This work was supported by the Affiliates Program of the Nanoscale Protoyping Laboratory

\bibliography{KMnAsO}

\end{document}